\documentclass[%
 reprint,
 amsmath,amssymb,longbibliography,
 aps,
]{revtex4-1}

\usepackage[english]{babel}
\usepackage{graphicx}% Include figure files
\usepackage{dcolumn}% Align table columns on decimal point
\usepackage{bm}% bold math
\usepackage{xcolor}
\usepackage{soul}
\begin{document}

%\preprint{APS/123-QED}

\title{Jamming Energy Landscape is Hierarchical and Ultrametric}% Force line breaks with \\
%\thanks{A footnote to the article title}

\author{R. C. Dennis}
\author{E. I. Corwin}
\affiliation{Department of Physics and Materials Science Institute, University of Oregon, Eugene, Oregon 97403, USA.}%\noaffiliation

\date{\today}% It is always \today, today,

\begin{abstract}
The free energy landscape of mean field marginal glasses is ultrametric. We demonstrate that this feature persists in finite three dimensional systems which are out of equilibrium by finding sets of minima which are nearby in configuration space. By calculating the distance between these nearby minima, we produce a small region of the distance metric. This metric exhibits a clear hierarchical structure and shows the signature of an ultrametric space. That such a hierarchy exists for the jamming energy landscape provides direct evidence for the existence of a marginal phase along the zero temperature jamming line.
\end{abstract}

\maketitle

The energy landscape surrounding a crystalline material clearly reflects the underlying crystal symmetries. Likewise, the energy landscape surrounding an amorphous material must reflect the replica symmetries underlying amorphous systems. The replica theory of glasses has shown that in the mean field limit, amorphous systems can exist in the liquid phase, the stable glass phase, or the marginal Gardner phase~\cite{gardner_spin_1985, mezard_spin_1987, castellani_spin-glass_2005, kurchan_exact_2013, charbonneau_fractal_2014, charbonneau_exact_2014, charbonneau_glass_2017, maiti_ergodicity_2018}. The energy landscape of the liquid phase is a single smooth basin, reflecting the unbroken replica symmetry of an ergodic phase. In the stable glass phase, this replica symmetry is broken and the landscape consists of many smooth basins separated by energy barriers~\cite{charbonneau_glass_2017}. However, within any individual basin, replica symmetry is still present. In the marginal Gardner phase, the replica symmetry is infinitely broken as each sub-basin is itself broken up into many sub-basins \textit{ad infinitum}~\cite{parisi_mean-field_2010, kurchan_exact_2013, charbonneau_exact_2014, rainone_following_2015, rainone_following_2016, urbani_shear_2017, biroli_liu-nagel_2018, scalliet_marginally_2019}. In the mean field framework, jamming is predicted to lie within the marginal Gardner phase \cite{rainone_following_2015, rainone_following_2016, biroli_breakdown_2016, urbani_shear_2017, scalliet_marginally_2019}. Indirect evidence for this phase in thermal systems has been observed in numerical simulations~\cite{berthier_growing_2016, scalliet_absence_2017, seoane_spin-glass-like_2018, hicks_gardner_2018, jin_stability-reversibility_2018, liao_hierarchical_2019, artiaco_exploratory_2019}, in two dimensional pseudo-thermal granular systems~\cite{seguin_experimental_2016}, and in thermal colloidal systems~\cite{hammond_seeing_2019}. The mean field result is applicable to low dimensional systems as evidenced by a recent result demonstrating through thermal exploration that the free energy landscape of quenched soft spheres has a hierarchical structure~\cite{artiaco_exploratory_2019}. Similarly, the free energy landscape of thermal disks at low temperatures has been observed to be hierarchical~\cite{liao_hierarchical_2019}. However, it is unknown how well this theory relates to physically relevant three dimensional athermal jammed packings~\cite{ohern_jamming_2003, majmudar_jamming_2007, charbonneau_universal_2012, morse_geometric_2014} for which not only are dynamics absent, but the system need not be created by an equilibrium process, and for which all behavior is solely determined by geometry.
%In this paper (at the end of the first paragraph)...
In this paper, we directly measure the Gardner phase in over-jammed systems by constructing the distance metric between nearby minima and characterizing its hierarchy and ultrametricity. We find that for a range of pressures, jammed systems are both hierarchical and ultrametric.  

\begin{figure}[]
\includegraphics[width=.475\textwidth]{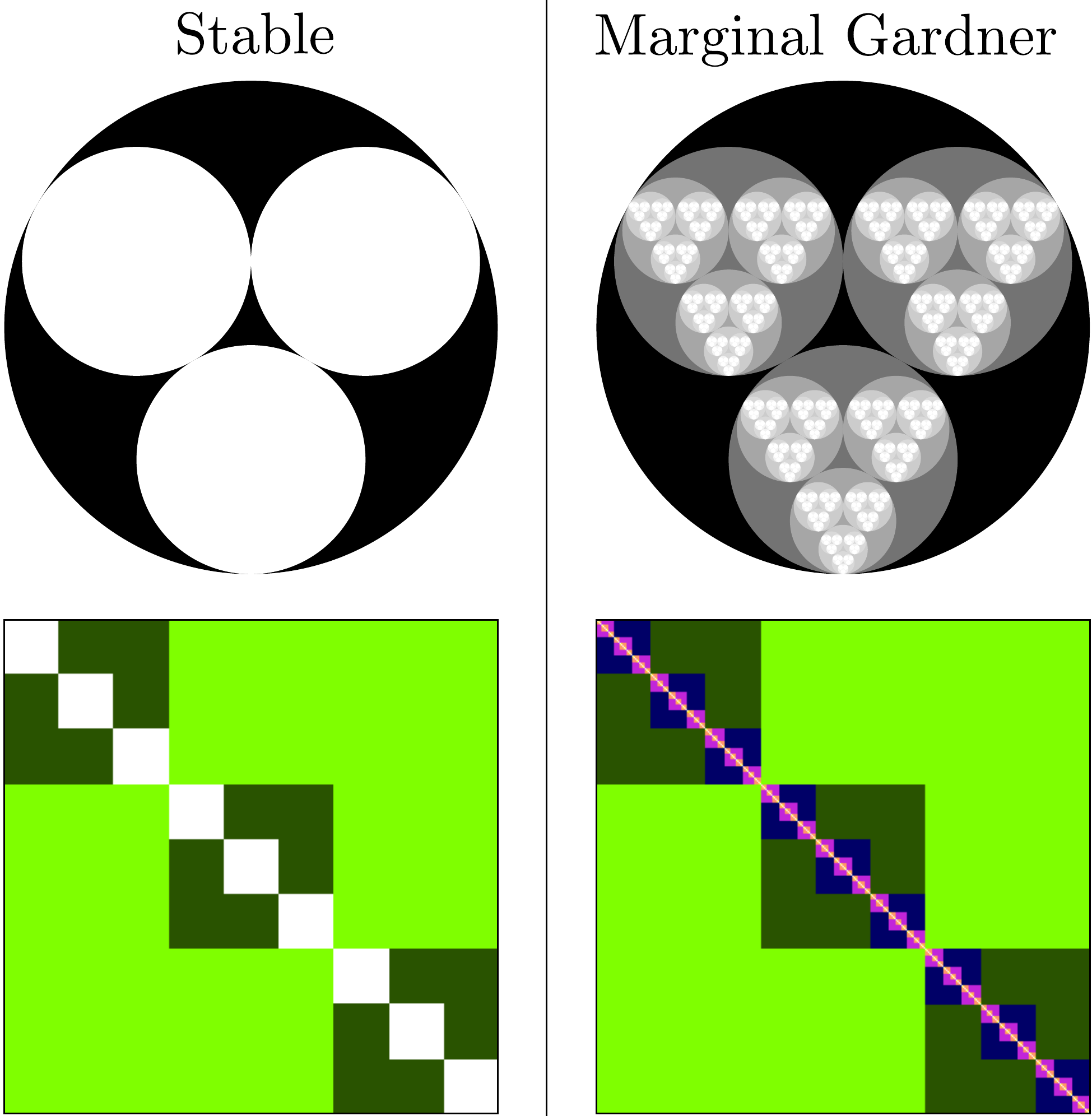}
\caption{Above: two dimensional schematic illustrations of the energy landscape present in the stable (1RSB) and marginal Gardner (fullRSB) phases, below: their respective metrics. The $ij$ entry in the metric describes the distance between minimum $i$ and minimum $j.$ The stable system has two levels of distinct infinitely long-lived free energy basins, shown as the set of circles contained within a larger circle. The metric for the stable phase likewise reflects this hierarchy, shown schematically below.  In the marginal Gardner system, every sub-basin has sub-basins forming a fractal energy landscape. The metric for such a landscape reflects marginality and is shown schematically below. Note that we depict each basin as having the same number of sub-basins, but marginal systems do not generally have this feature.}\label{RSBFigure}
\end{figure}

%Introduce the concept of a metric
As illustrated in Figure~\ref{RSBFigure}, the single replica symmetry breaking (1RSB) solution reflects the fact that a stable glass phase is characterized by {\em distinct}, infinitely long-lived energy basins. The solution with infinitely many distinct basins-within-basins representing the marginal Gardner phase is called the fullRSB solution~\cite{ gardner_spin_1985, marinari_replica_2000, charbonneau_fractal_2014}. The hierarchical structure of a marginal Gardner phase results in minima forming a tree-like structure in phase space for which minima within a given sub-basin will all be much closer to one another than they will be to minima within any other sub-basin~\cite{castellani_spin-glass_2005}. This feature is codified by the ultrametric inequality~\cite{murtagh_ultrametricity_2004, schneider_ultrametricity_2009} which states that the distance $d$ between any three configurations $a, b,$ and $c$ must satisfy
\begin{eqnarray}
d\left(a,c\right)\leq \max{\left[d\left(a,b\right),d\left(b,c\right)\right]}.
\end{eqnarray}

We construct jammed packings of $N$ monodisperse soft spheres interacting through a harmonic contact potential in three dimensions using the FIRE algorithm~\cite{bitzek_structural_2006} as implemented by the pyCudaPacking software~\cite{morse_geometric_2014, charbonneau_universal_2016, morse_echoes_2017}. In order to unambiguously distinguish nearby minima in the energy landscape, all calculations are done with quad precision floating point numbers and minimization is only halted once the maximum unbalanced force on any particle is less than $10^{-20}$ in natural units. Systems are created in a cube of side length 1 with periodic boundary conditions and at a large initial packing fraction $\phi = 0.8.$ These packings are then brought to a specified pressure~\cite{allen_computer_2017} through an iterative process exploiting the known scaling between packing fraction and pressure for over-jammed systems \cite{charbonneau_jamming_2015}.

In sufficiently small systems ($N\sim10$ in two dimensions), one can sample the entire energy landscape, enumerate all minima, and use these minima to construct the metric for the landscape~\cite{gao_enumeration_2007, ashwin_calculations_2012}. However, this quickly becomes intractable as the number of minima increases exponentially with increasing $N.$ Choosing energy minima at random results in a small uncorrelated sample which will trivially not reveal any hierarchical structure as it is extraordinarily unlikely that two minima will be a part of the same deep sub-basin~\cite{xu_direct_2011, berthier_microscopic_2011}. Instead, we search for correlated samples with a small number of minima which are close together in configuration space and thus have the power to reveal any existing hierarchy.

To explore behavior as a function of distance to jamming, we create initial systems at logarithmically spaced pressures, $p,$ running from $10^{-1}$ down to $10^{-5.5}$ in natural units. Given a system at a specified pressure, we explore the nearby minima that characterize the local energy landscape by repeatedly perturbing the initial conditions of the original minimum and re-minimizing. Each perturbation is chosen randomly from a Gaussian distribution and amounts to moving each particle a random distance in a random direction. Due to the random nature of the perturbation, there will be a small component of global particle translation.  To remove this we subtract off the global translation when calculating $\varepsilon$, the magnitude of the perturbation. Further, this magnitude is normalized by the typical interparticle spacing, $N^{-1/3},$ to remove the trivial dependence on the number of particles in the system in a way which is independent of the system's packing fraction.

Depending on the initial pressure, many to most nearby perturbed systems will return to the original configuration. To adequately sample the nearby landscape, we continue to perturb the original minimum until we have found 500 distinct minima (with the exception of the data presented in Figure~\ref{fig:ultrametricity} for which 5000 minima were found).

\begin{figure}[]
\includegraphics[width=0.475\textwidth]{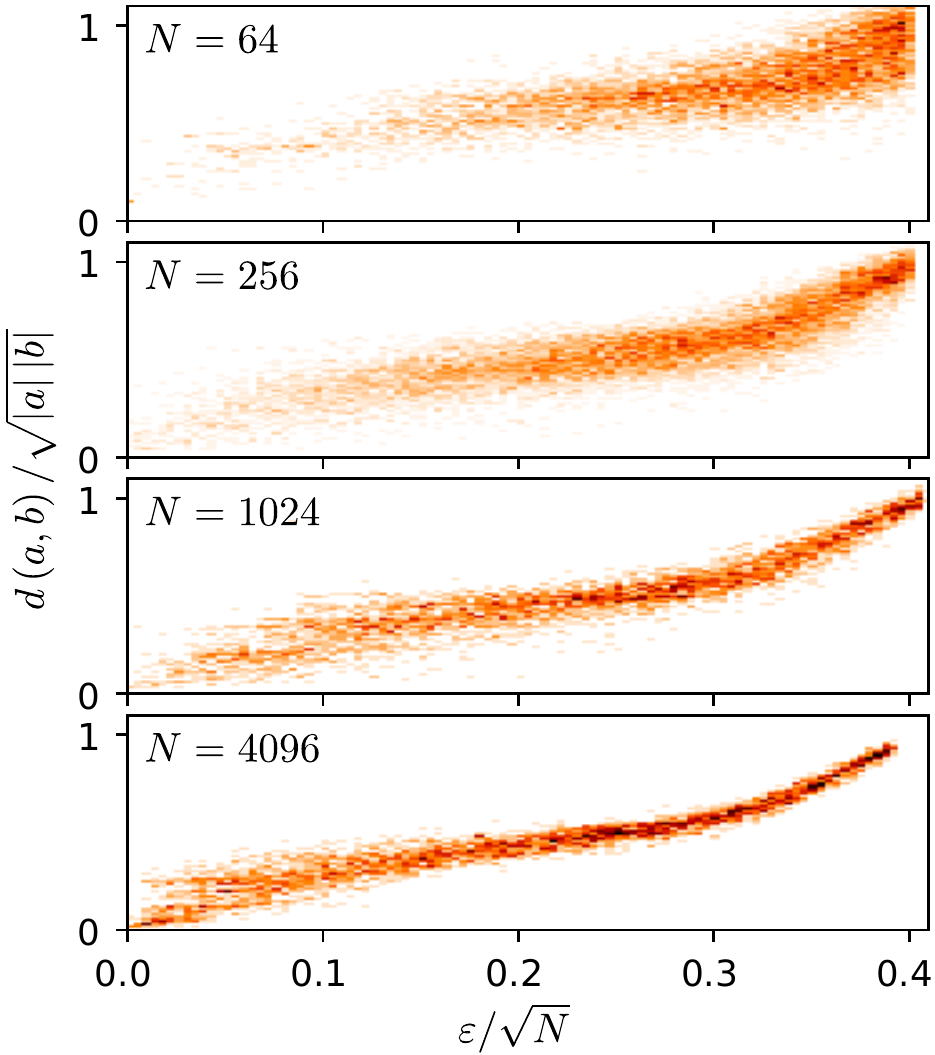}
\caption{Two dimensional histograms of the distribution of normalized metric distance to the original packing as a function of the size of the perturbation. The original packing is perturbed by a Gaussian random vector with length $\varepsilon$ and then minimized. The distance between the original minimum and this newly discovered minimum, $d\left(a, b\right),$ is found. This distance is normalized by $\sqrt{\left|a\right|\left|b\right|}$ where the absolute value of a system $\left|a\right|,$ is defined as $d\left(a,0\right)$ and $0$ is the contact network containing all zeros. From the top, plots for packings with 64, 256, 1024, and 4096 particles at pressure $p=10^{-3}.$ We see that these curves all take a similar functional form and have a normalized metric distance of about $1$ at $0.4\sqrt{N}$ which is thus a natural value for $\varepsilon_{\textrm{max}}.$}\label{epsilonVOverlap}
\end{figure}

Finding the metric for nearby minima using the perturbation technique requires choosing a length scale for the perturbation. A perturbation which is too small will frequently lead back to the original minimum. A perturbation which is too large will result in minima which do not fall within the same top-level super-basin and are not sufficiently nearby in configuration space to properly probe the hierarchical structure of the landscape. Because the configuration space is $Nd$ dimensional, sampling a small spherical volume of the space biases points to the surface of the sphere. Instead, to better sample nearby minima, the length of the perturbation $\varepsilon$ is chosen from a uniform distribution between 0 and $\varepsilon_\textrm{max}.$ 

\begin{figure*}[]
\includegraphics[width=1\textwidth]{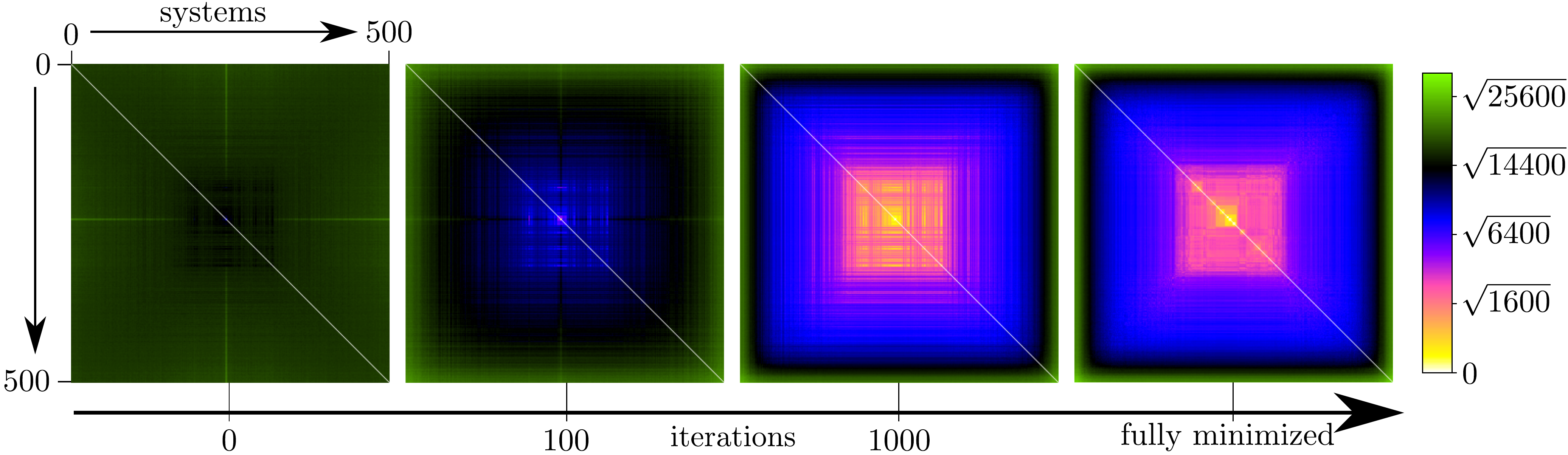}
\caption{Evolution of hierarchy with minimization. $500$ configurations with $N=4096$ are prepared by perturbing a random minimum at a pressure $10^{-3}.$  The metric distance between every pair of configurations, as given in equation~\ref{cvDef}, is shown for 0, 100, and 1000 minimization iterations as well as for fully minimized systems. The color scale reflects the metric distance is labelled by square rooted numbers, reflecting the fact that the metric distance is roughly the square root of the number of changed contacts between two systems for $d(a,b) < \sqrt{N}$.}\label{iterations}
\end{figure*}

Figure~\ref{epsilonVOverlap} shows the magnitude of the initial perturbation, scaled by $\sqrt{N}$, plotted against the resulting normalized distances between the original system and the minimized perturbed system. A scaled distance of one means that the number of stable contacts that differ between two systems is comparable to the number of stable contacts present within each system. Systems that are greater than this distance bear no more structural relationship and are thus in different top-level basins, making this a natural cutoff for exploring the hierarchical structure of the local energy landscape. The relationship between distance and initial perturbation becomes sharper with increasing $N$ and does not depend strongly on pressure. Exploiting this empirical relationship, we set $\varepsilon_\textrm{max}=0.4\sqrt{N}.$

Given a set of nearby minima, we construct the metric $d$ by calculating the distance between every pair of minima. To avoid the ambiguity introduced by rattlers and by global drifts, we define the distance based on the stable contact vector network within each system. The stable contact vector between particle $i$ and particle $j$ for configuration $a$ is denoted as $\vec{C}_a^{ij}.$ If two particles are not in contact, the contact vector between them is taken to be $\vec{0}.$  The distance between two systems $a$ and $b$ is
\begin{eqnarray}
\label{cvDef}
d \left( a,b \right) \equiv \frac{1}{\langle\sigma \rangle} \sqrt{\sum_{ij} \left( \vec{C}_a^{ij}-\vec{C}_b^{ij} \right)^2}
\end{eqnarray}
where $\sigma$ is the diameter of a particle.
This metric has the convenient property that $d(a,b)$ will be approximately equal to the square root of the number of contacts that differ between the two minima for $d(a,b) < \sqrt{N}$.

For any set of elements with a metric, one can construct a new ultrametric by changing the pairwise distances. There exists a family of ultrametrics for which every distance is smaller than that found in the original metric. Of these, the ultrametric that is closest to the original metric is called the subdominant ultrametric, $d^<,$ and can be constructed from the original metric using a minimum spanning tree~\cite{kruskal_shortest_1956, rammal_degree_1985} as detailed in the supplementary information. We characterize the generalized distance between the subdominant ultrametric and the original metric as 
\begin{eqnarray}
\mathcal{D} \equiv \sqrt{\left\langle\left(d(a,b)-d^{<}(a,b) \right)^2\right\rangle}
\label{dou}
\end{eqnarray}
where the angle brackets denote an average taken over every pair of $a$ and $b.$ $\mathcal{D} = 0$ indicates a precisely ultrametric system.

\textit{Development of hierarchy upon minimization --} Figure~\ref{iterations} shows the evolution of the metric between distinct nearby minima of $N=4096$ particles as a function of iterations of the minimization protocol. These 500 minima are all initially created by the above perturbation process around an arbitrarily chosen initial minimum. The simple nature of this random perturbation is revealed in the first panel which shows every system is initially nearly equidistant (shown in black and dark green). After 100 iterations (second panel) of minimization, the structure of a basin (shown in black and blue) begins to appear as some systems relax towards the initial minimum by reforming contacts; meanwhile others relax away by forming different contacts and fall into distinct super-basins (shown in lighter green). After 1000 iterations (third panel), the hierarchical structure begins to appear but only becomes fully realized once systems are fully minimized (final panel). The metrics are all sorted using the single link clustering algorithm~\cite{murtagh_survey_1983} on the subdominant ultrametric of the fully minimized systems.

\begin{figure*}[]
\includegraphics[width=1\textwidth]{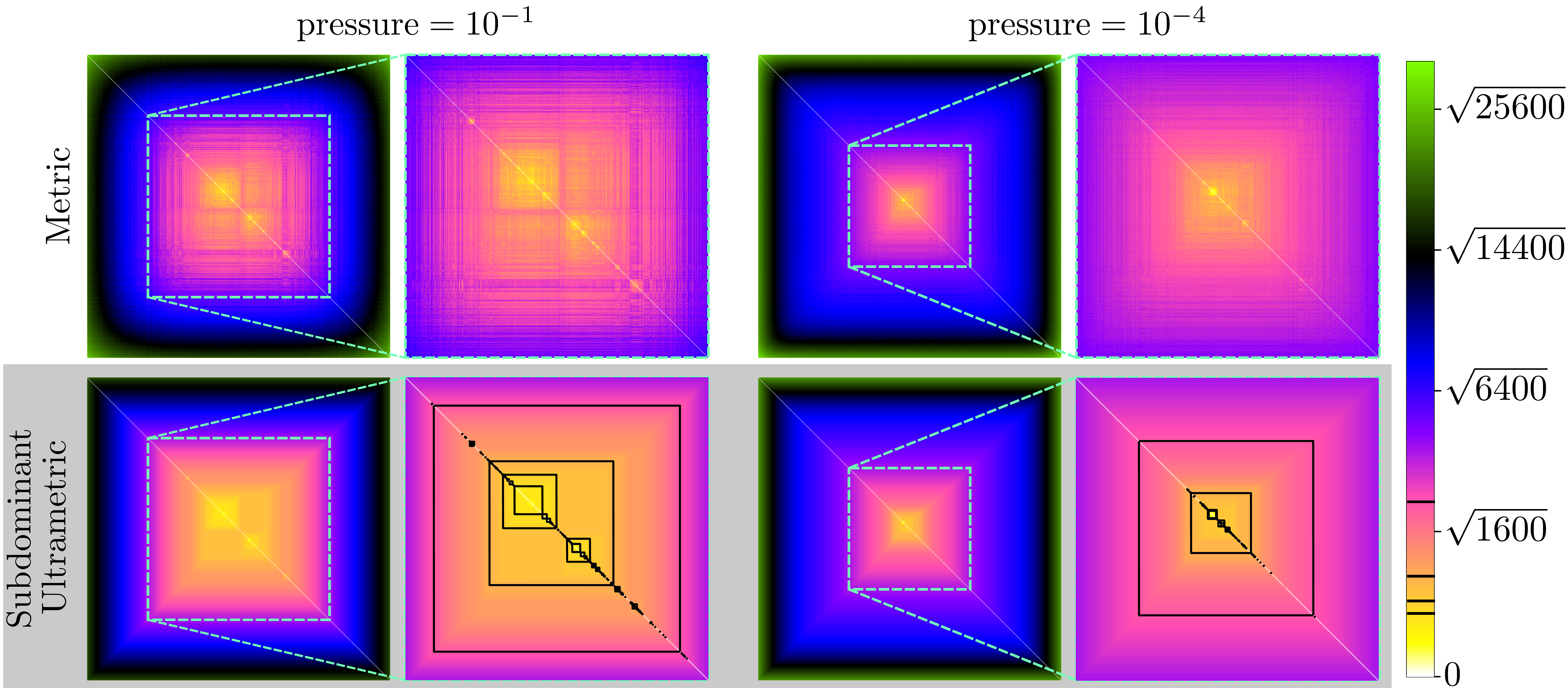}
\caption{Metrics (top) and corresponding subdominant ultrametrics (bottom) as a function of pressure constructed from 5000 systems with $N=4096$ particles.  Next to each metric and ultrametric is a blowup of the region for which the subdominant ultrametric distance is less than $\sqrt{4096}$ which amounts to approximately 1 contact per particle. Contours of the subdominant ultrametric are overlayed to highlight the hierarchy and their values are shown on the color bar. The color scheme is the same as in Figure~\ref{iterations}.}\label{fig:ultrametricity}
\end{figure*}

\textit{The hierarchical structure at different pressures --} We plot the metric and corresponding subdominant ultrametric for minima of $N=4096$ particles far from jamming, $p = 10^{-1},$ and those closer to jamming, $p = 10^{-4},$ in Figure~\ref{fig:ultrametricity}. These metrics are each constructed from 5000 distinct nearby minima. As jamming is approached, we observe the metric to become more similar to the subdominant ultrametric and we see that ever fewer minima fall into the same sub-basins. Visually, systems at a low pressure have a metric that is closer to the subdominant ultrametric than do those at high pressure. This can be observed in the quality of the color scale matching and the sharpness of the boxes corresponding to sub-basins. For the high pressure metric, three-fifths of all systems differ from one another by less than one contact per particle whereas at low pressure about two-fifths of the systems differ by less than this amount. Once perturbed, the positions of particles for low pressure systems do not need to change as much before finding a new minimum. As the pressure is decreased, the number of nearby minima explodes leading to a shrinking of the region that can be densely sampled. Both of these results arise from the increasingly rough and hierarchical energy landscape upon the approach to jamming.

\begin{figure}[]
\includegraphics[width=0.475\textwidth]{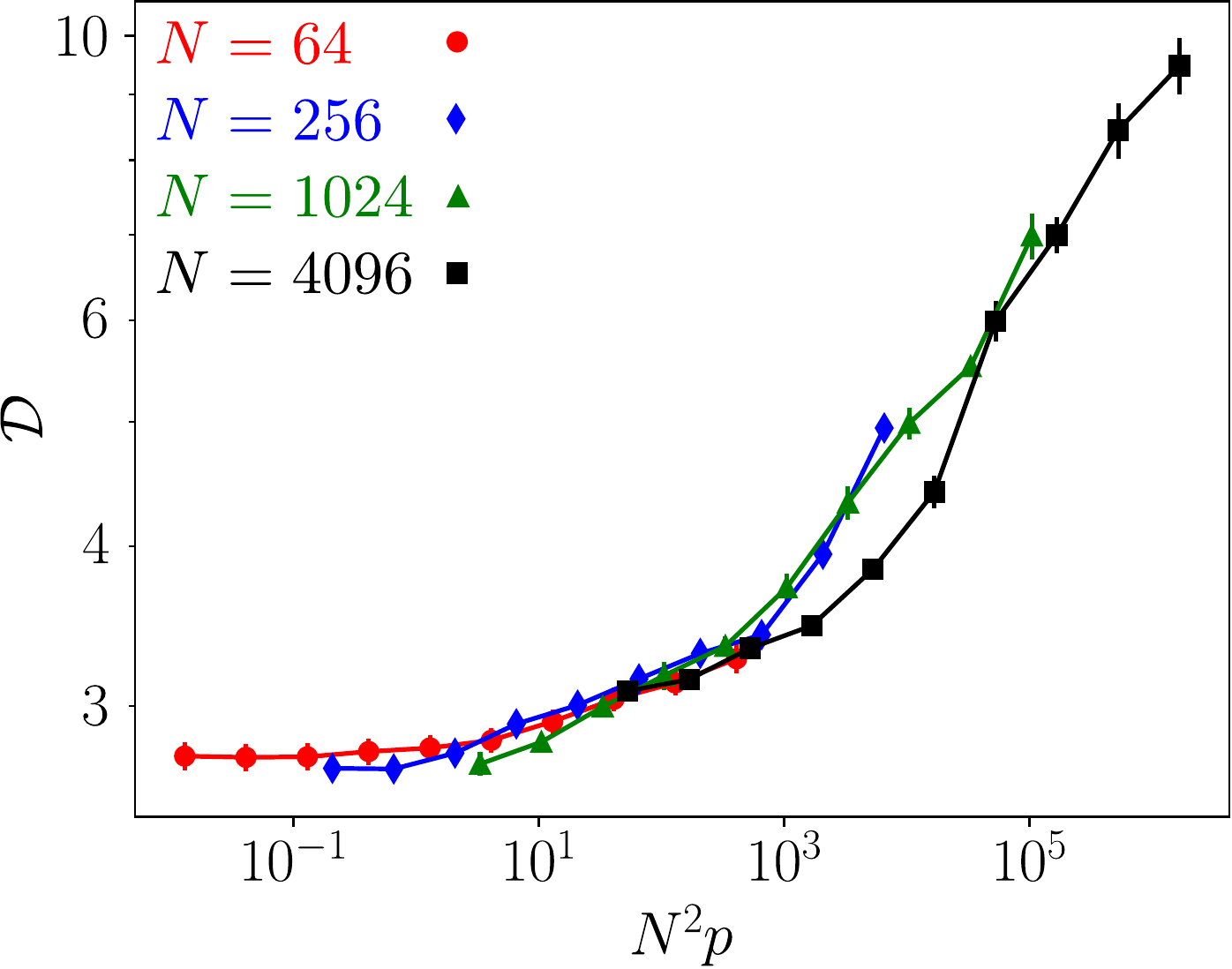}
\caption{
The generalized distance between the subdominant ultrametric and the original metric, $\mathcal{D},$ as a function of pressure and system size. The number of systems over which each point is averaged is chosen such that the standard error bars fall below a threshold. Systems of different sizes fall on a master curve.}\label{fig:DOU}
\end{figure}

We quantify the qualitative result of increasing ultrametricity with decreasing pressure in Figure~\ref{fig:DOU} by plotting $\mathcal{D}$ as a function of scaled pressure, $N^2p,$ which can be interpreted as the distance to jamming~\cite{goodrich_finite-size_2012}. We see that for all system sizes $\mathcal{D}$ collapses onto a master curve which achieves a plateau value of about 2.7 as $N^2p$ goes to zero. This means that on average the distance between any pair of minima will be bigger than the distance needed for ultrametricity by about 2.7. However, the distance between any pair of minima itself scales with $\sqrt{N}$ so this \textit{fractional} excess of distance will tend to zero as $N$ becomes large. Therefore in the thermodynamic limit the metric becomes precisely ultrametric for all of the pressures explored.

\textit{Conclusions --} The structure of the distance metric between minima provides the first evidence that the energy landscape of over-jammed three dimensional configurations becomes hierarchical and ultrametric in the thermodynamic limit for all pressures sampled. In this limit, the marginal Gardner phase arises as strictly a consequence of geometry with no recourse to thermal fluctuations. It is far from clear that this hierarchy and ultrametricity arises for such low-dimensional configurations, especially with finite numbers of particles. This result points to the universality of the marginal Gardner phase within amorphous materials as it has now been measured within athermal materials in addition to the already known thermal~\cite{liao_hierarchical_2019, artiaco_exploratory_2019} and mean-field limits~\cite{mezard_spin_1987}. By exploring the energy landscape at zero temperature and never with any sense of thermal exploration, we have sampled a spatially localized region of phase space. Our results demonstrate that the Gardner phenomenology is not just restricted to the easily accessible regions of configuration space that are seen in thermal materials, but is instead present everywhere.

This research demonstrates that Gardner physics can be observed in athermal out-of-equilibrium systems.  Furthermore, that this result can be seen in an athermal system demonstrates that the Gardner transition controls not only the free energy landscape but also the underlying energy landscape.  As such, Gardner physics should be amenable to experimental tests which need not rely on thermal systems. This innovation marks a significant step forward in fully understanding glasses and jammed materials as we unify the concept of marginality in amorphous systems.

\textit{Acknowledgments --} We thank Horst-Holger Boltz, Peter Morse, Sid Nagel, and Camille Scalliet for useful discussions. This work benefited from access to the University of Oregon high performance computer, Talapas. This work was supported by National Science Foundation (NSF) Career Award DMR-1255370 and the Simons Foundation No. 454939.

\bibliography{hierarchicalJamming}

\widetext
\clearpage

\begin{center}
\textrm{\large Supplementary Materials}
\end{center}
\section*{The Subdominant Ultrametric}
We present here a simple outline of the algorithm for creating the subdominant ultrametric and provide an intuitive explanation for how it works.  Given a metric, $d,$ the corresponding subdominant ultrametric, $d^{<},$ can be found with the following algorithm~\cite{rammal_degree_1985}:
\begin{enumerate}
\item The metric, $d,$ is a symmetric matrix of pairwise distances between minima. This can be reinterpreted as an edge-weighted graph where the nodes are the minima and the edge weights are the distances between minima.
\item We compute the minimum spanning tree of this graph, which is simply the network with the minimum possible total edge weight (sum of distance values) which connects every node into a single tree. The minimum spanning tree is unique~\cite{kruskal_shortest_1956} and has the property that every pair of nodes has only one path connecting them.
\item The subdominant ultrametric, $d^{<},$ is created as a symmetric matrix with entries $d^{<}_{ij}$ determined by the maximum edge weight in the path from node $i$ to node $j$ in the minimum spanning tree.
\end{enumerate}
The hierarchical nature of $d^<$ derives from that of the minimum spanning tree. The maximum condition ensures that the resulting metric will be ultrametric as it is a direct enforcement of the ultrametric inequality (Equation~1 of the text). Given any triplet of minima the constructed subdominant ultrametric $d^<$ will produce a triangle with one short side and two equal long sides. This is an equivalent definition of the ultrametric inequality given in the manuscript. This explanation demonstrates why this algorithm returns an ultrametric which always contains distances that are less than or equal to the corresponding metric entries.  However, the proof that this is the largest possible ultrametric to satisfy this criterion is less intuitively obvious and can be found in the original reference~\cite{rammal_degree_1985}.
\subsection*{Uniqueness of the Subdominant Ultrametric}
If all of the edge-weights in our metric are distinct, the minimum spanning tree will be unique~\cite{kruskal_shortest_1956}. Our metrics come from amorphous systems with unique edge-weights and unique minimum spanning trees. Additionally, the subdominant ultrametric is always unique whether or not this condition is met because while degenerate edge-weights results in multiple minimum spanning trees, the maximum edge-weight along the path between every pair of nodes will be the same~\cite{rammal_degree_1985}.
\end{document}